\documentclass[prl,twocolumn,showpacs,amsmath,amssymb]{revtex4-1}

\usepackage{graphicx}
\usepackage{amsfonts}
\usepackage{amsmath}
\usepackage{amssymb}
\usepackage{color}
\usepackage{bm}
\usepackage{enumerate}
\usepackage{color}
\graphicspath{figures} 
\usepackage{amsfonts, amsmath, amsthm, amssymb} 
\usepackage{mathtools}
\usepackage{array}
\usepackage[colorlinks,bookmarks=false,citecolor=blue,linkcolor=red,urlcolor=blue]{hyperref}

\usepackage[colorlinks,bookmarks=false,citecolor=blue,linkcolor=red,urlcolor=blue]{hyperref}
\usepackage[usenames,dvipsnames,svgnames,table]{xcolor}

\usepackage[T1]{fontenc}          
\usepackage{fancybox}		   
\usepackage{makeidx} 

\usepackage{siunitx}
\usepackage{amsmath}

\begin{document}                  

\title{Measurement of the Casimir force in a gas and in a liquid}     

\author{Anne Le Cunuder $^{1,*}$, Artyom Petrosyan $^1$, Georges Palasantzas $^2$, Vitaly Svetovoy $^{2,3}$ and Sergio Ciliberto $^1$}        

\affiliation{$^1$Laboratoire de Physique, CNRS UMR5672,  Universit\'e de Lyon, \'Ecole Normale Sup\'erieure, \\
46 All\'ee d'Italie, 69364 Lyon, France} 
\affiliation{$^2$ Faculty of Sciences and Engineering,  University of Groningen,\\ Nijenborg 4, 9747 AG Groningen, The Netherlands}
\affiliation{$^3$ A. N. Frumkin Institute of Physical Chemistry and Electrochemistry, Russian Academy of Sciences, Leninsky prospect 31 bld. 4, 119071 Moscow, Russia}

\begin{abstract}

We present detailed measurements of the Casimir-Lifshitz force between two gold surfaces, performed for the first time in both gas (nitrogen) and liquid (ethanol) environments with the same apparatus and on the same spot of the sample. Furthermore, we study the role of double-layer forces in the liquid, and we show that these electrostatic effects are important. The latter contributions are subtracted to recover the genuine Casimir force, and the experimental results are compared with calculations using Lifshitz theory. Our measurements demonstrate that carefully accounting for the actual optical properties of the surfaces is necessary for an accurate comparison with the Lifshitz theory predictions at distances smaller than 200nm.

\end{abstract} 
		
\maketitle                       

\textit{Introduction.--} As devices enter the submicron range, Casimir forces \cite{Genet,Casimir:1948dh,Lifshitz:1956zz,PhysRevLett.78.5,Iannuzzi4019,Chen:07,PhysRevLett.107.090403,PhysRevLett.103.040402,PhysRevA.82.010101,PhysRevB.77.035439,PhysRevB.85.045436,PhysRevD.75.077101} between neutral bodies at close proximity become increasingly important. As Casimir first understood in 1948 \cite{Casimir:1948dh}, these forces are due to the confinement of quantum fluctuations of the electromagnetic (EM) field. Indeed Casimir proved that when two parallel, perfectly reflecting plates, are introduced in vacuum, they impose, on the EM field, boundary conditions which select only the fluctuations compatible with them. As a result, an attractive force between the plates is produced, which depends only on fundamental constants, on the distance $d$ between the surfaces and on their area $A$:
\begin{equation}
F_c(d)=-\frac{\pi^2 A \hbar c}{240 \  d^4}
\end{equation}
with $\hbar$ the Planck constant and $c$ the speed of light. Following Casimir's calculation \cite{Casimir:1948dh}, Lifshitz and co-workers in the 50's \cite{Lifshitz:1956zz} considered the more general case of real dielectric plates by exploiting the fluctuation-dissipation theorem, which relates the dissipative properties of the plates  and EM fluctuations at equilibrium. Furthermore, for real surfaces, roughness and material optical properties can strongly alter the Casimir force \cite{SVETOVOY20151,Roughness_2012}.

\begin{figure}
\centering
\includegraphics[width=1\linewidth]{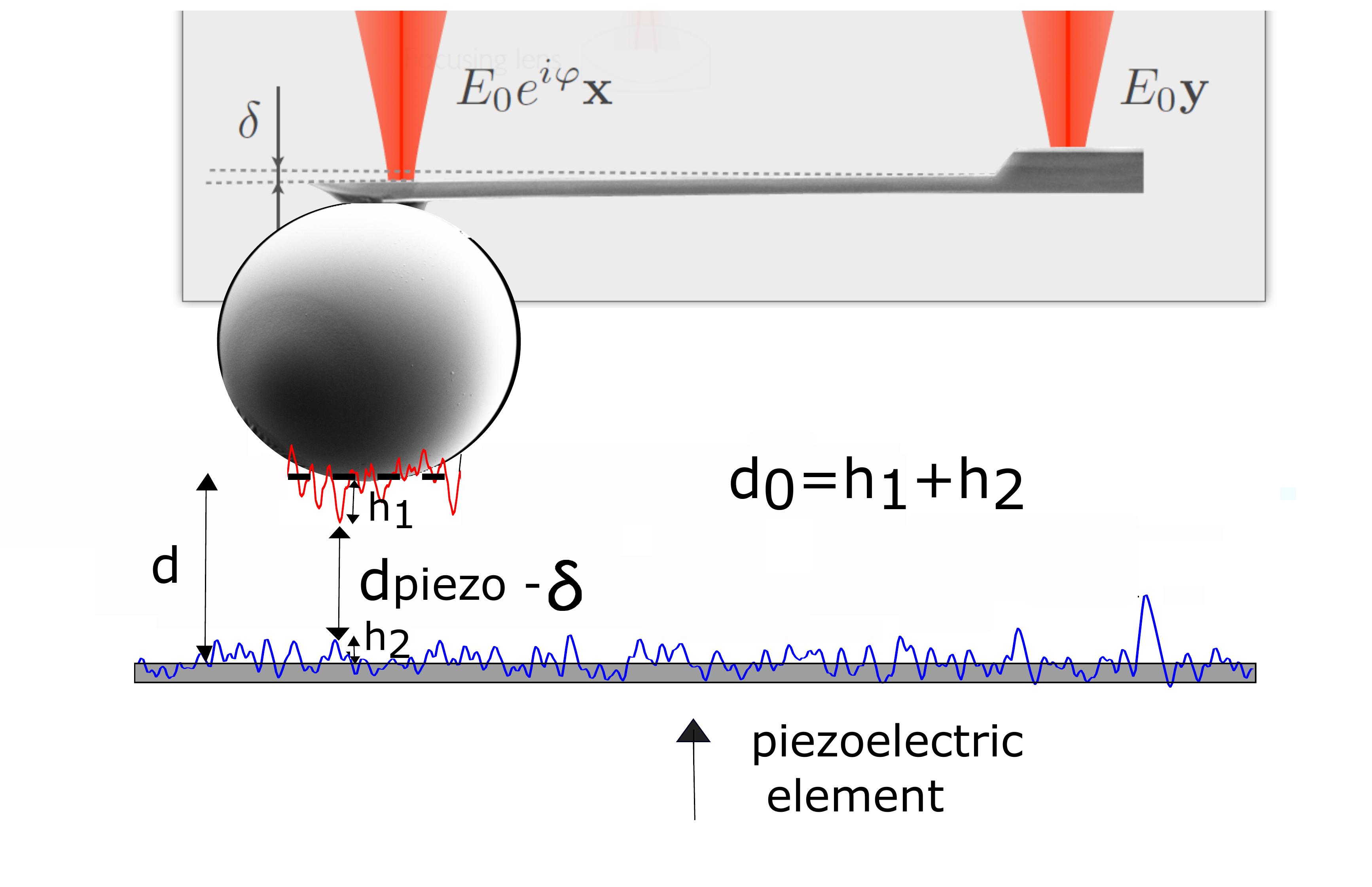}
\caption{\textbf{Experimental setup.} A gold-coated polystyrene bead is glued on the cantilever tip which measures the sphere-plane interaction force at distance $d$. The deflection of the cantilever is detected by an interferometric technique: two lasers beams, orthogonally polarized, are focused on the cantilever, the reference one is reflected by the static base and the second one by the cantilever free end. When the cantilever is bended the optical path difference $\delta$ between the two beams is measured through an interferometer \cite{bellon:tel-00541336}.}
\label{fig:interferometer_principle} 
\end{figure}

Lifshitz formalism describes the Casimir force in a general case, where the medium between the plates need not be vaccum. According to this formalism, the force can be tuned from attractive to repulsive with a suitable choice of the interacting materials. These predictions boosted Casimir experiments to test the possibility of repulsive forces \cite{munday2009measured}. In liquids, the determination of the Casimir force is more complex than in a gas because of the presence of additionnal effects, as e.g. the Debye screening. The Casimir-Lifshitz force has been measured between two gold surfaces immersed in ethanol \cite{PhysRevA.75.060102}; in this experiment, electrostatic forces are found to be negligible as sodium iodide (NaI) was added to ethanol, decreasing the Debye screening length. However, the role of electrostatic forces and their screening by the Debye layer are important and one has to consider carefully their contributions during force measurements in liquids \cite{PhysRevA.77.036102,PhysRevA.77.036103}. In order to clarify the interplay of the Casimir force and additionnal effects in liquids, we have performed measurements of the Casimir force in a nitrogen atmosphere in the first place, and then, using the same system and sample, in ethanol. The contact area is the same in both measurements. We observe that electrostatic forces, screened over the Debye length, are of the same magnitude as the Casimir force, in the 50-200nm distance range. After subtracting the electrostatic force, we obtain a Casimir force in quantitative agreement with Lifshitz theory \cite{Lifshitz:1956zz}. Furthermore, the accuracy of our measurement allows us to highlight the importance of accurately characterizing the optical properties of the samples before any meaningful comparison with theory.

\textit{Experimental setup.--}\label{sec:experiment} We use an atomic force microscope (AFM) to measure the  Casimir force between metallic surfaces. In order to measure the force with a good accuracy, the cantilever displacement is measured with a home-made quadrature phase interferometer \cite{bellon:tel-00541336}, whose operating principle is sketched on Fig.~\ref{fig:interferometer_principle}.

 	The experiment is performed in a sphere-plane geometry to avoid the need to maintain two flat plates perfectly parallel. Thus, a polystyrene sphere of radius  $R=({75.00 \pm 0.25} ) \si{\micro\meter}$ (Sigma-Aldrich) is mounted on the tip of the cantilever with a conductive glue and then the whole probe is coated by a gold film whose thickness is about ${100}\si{\nano\meter}$. The plates have been gold coated using cathodic sputtering by \href{http://www.acm-pvd.com/en/}{ACM}, at the \href{http://lma.in2p3.fr/}{LMA-CNRS}.
The diameter of the sphere has been determined from Scanning Electron Microscopy. 
 We use a cantilever (size ${500}\si{\micro\meter} \times {30}\si{\micro\meter} \times {2.7}\si{\micro\meter}$, NanoAndMore) of  stiffness $\kappa={0.57 \pm 0.03}\si{\newton / \meter}$. The precise value of $\kappa$ is determined  using equipartition, ie. $<\delta^2>=\frac{k_B T}{\kappa}$, where $k_B$ is the Boltzmann constant and $T$ the temperature. The resonance frequency of the sphere-cantilever ensemble  is $f_o=2271 \si{Hz}$ in vacuum.

The sphere faces a glass flat plate which is coated by a gold film of a thickness of about $100\si{\nano\meter}$ \cite{SuppMat}. According to \cite{Lisanti11989}, the layer is sufficiently thick to be considered as a bulk-like film. The plate is mounted on a piezo actuator (PZ$38$, Piezojena) which allows us to control the plane-sphere distance. During the experiment, the plate is moved continuously towards the sphere and the induced deflexion of the cantilever is detected by the interferometer.

In air, because of water vapor, the capillary force far exceeds the Casimir force. Therefore, our measurement in a gas was performed after filling our cell with nitrogen.

\textit{Calibrations.--} \label{sec:add} The total force between the surfaces is the sum of the Casimir force $F_\text{cas}(d)$ and additionnal contributions: 

\begin{equation}
F_\text{total}=F_\text{cas}(d)+F_\text{el}(d)+F_{H}(d,v) ~.
\label{eq:FT}
\end{equation}

Electrostatic forces $F_\text{el}(d)$ are due to a potential difference between surfaces, owing to differences between the work functions of the materials used, and the possible presence of trapped charges \cite{PhysRevLett.90.160403}. Hydrodynamic forces $F_{H}(d,v)$ are due to the motion of the fluid during the approach of the plate towards the sphere, and depend on their relative velocity $v$ \cite{BRENNER1961242}. These hydrodynamic effects are negligible in a nitrogen atmosphere, where the viscosity is $\gamma={1.76\ 10^{-6}}\si{\pascal \second}$, but have to be considered in ethanol where the viscosity is $1000$ times higher ($\gamma={1.2 \ 10^{-3}}\si{\pascal \second}$).

There are two main requirements for a precise determination of the Casimir force. Firstly, additionnal forces must be measured with accuracy and subtracted from the total measured force. Secondly, because the force has a strong dependance on the distance between surfaces, an independant measure of the distance is necessary, which becomes difficult  when the separation approaches nanometer scales. The difficulty originates principally from surface roughness: when the two surfaces come into contact, the highest asperities of each surface touch each other and the surfaces are still separated by a distance upon contact $d_0$ \cite{PhysRevB_distance}.

The piezo actuator includes a position sensor which gives us the displacement of the plate: $d_\text{piezo}$. We define the origin of $d_\text{piezo}$ as the position of contact of the highest peak of the sphere roughness with the surface of the plate, as the sphere is much rougher than the plate. The effective separation distance which appears in the expression of the force can be written as (see Fig.~\ref{fig:interferometer_principle}):

\begin{equation}
d=d_\text{piezo}+d_0-\delta
\label{distance_effective}
\end{equation}
where $d_0$ is the distance upon contact due to surface roughness and $\delta$  is an additional correction which results from the static deflexion of the cantilever in response to the total force $F_\text{total}$.

We determined the separation upon contact $d_0$ from hydrodynamic calibration, performed in ethanol. Immediatly after measuring the Casimir force in nitrogen, we injected carefully ethanol into the cell, and we performed calibrations and measurements of the Casimir force in ethanol. As the horizontal drift of our system is negligible, the contact area and the separation distance upon contact $d_0$ are the same in each measurement. This assumption is further justified a posteriori: our experimental curves all superimpose on top of each other and on top of the theoretical curve after shifting the distance by the same value of $d_0$. The hydrodynamic calibration is presented in next section, while topographic analysis is presented in \cite{SuppMat}. The value of $d_0$ obtained from hydrodynamic calibration is comparable with the value obtained from roughness analysis.

 \begin{figure*}[t]
\includegraphics[width=0.8\linewidth]{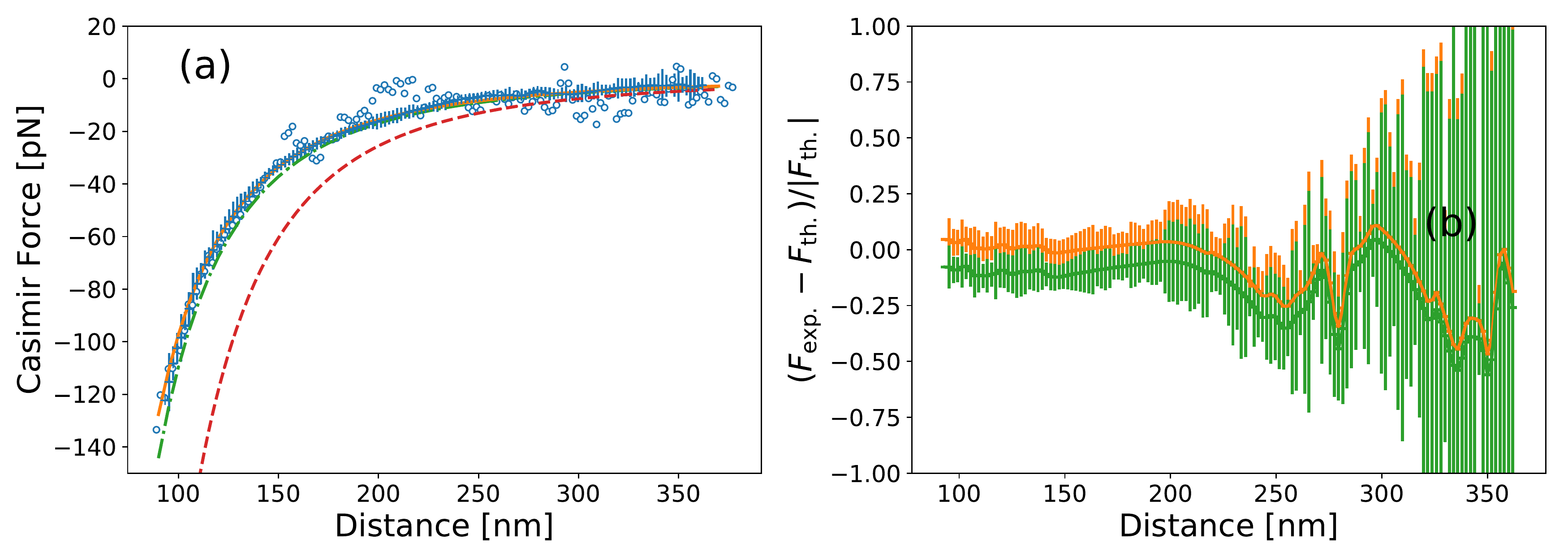}
\caption{\textbf{Measurement of the Casimir force between two Au-surfaces in a nitrogen atmosphere}. (a) Blue points correspond to the mean measured force. Blue circles correspond to a single measurement of the Casimir force. Orange solid curve corresponds to the Lifshitz theory in which the dielectric function $\epsilon$ is evaluated from measured optical data of a real gold film \cite{PhysRevB_optical}. For the green dashed-dotted line, $\epsilon$ is evaluated using the handbook optical data \cite{palik1998handbook}. Red dashed line corresponds to the theory in the case of ideal conductors. (b) Relative difference between the theoretical and experimental Casimir forces. Same colors as in (a). The error bars include the statistical and the systematic errors due to uncertainties on the separation upon contact {$d_0=(31 \pm 2)\si{\nano\meter}$ (using the hydrodynamic estimation)}, on the stiffness $\kappa=(0.57 \pm 0.03) \si{\newton / \meter}$ and on the diameter of the sphere $D=(150.0 \pm 0.5)\si{\micro\meter}$.}
\label{force_casimir_nitrogen}
\end{figure*}

\textit{Hydrodynamic calibration in ethanol.--} \label{sec:hydrodynamics}
The theoretical expression of the hydrodynamic force, for non-slip boundary conditions, is given by \cite{BRENNER1961242}:

 \begin{equation}
 F_H= - \frac{6 \pi \eta R^2}{d} v
\label{Taylor}
 \end{equation}
 where $\eta$ is the fluid viscosity, $R$ is the radius of the sphere, and $v=\frac{\partial d}{\partial t}$ is the relative velocity between the plate and the sphere. Indeed, in our case, the slippage can be neglected as the mean free path is in the order of inter-molecular distances (e.g a few Angstroms), and negligible in comparison with the roughness of the surface (e.g a few tens of nanometers) \cite{Physicsoffluids}

As is clear from Eq.~\eqref{eq:FT}, among the different forces occuring between the surfaces, the hydrodynamic force is the only one which depends on velocity. Thus we performed two force measurements, moving continuously the plate towards the sphere: a first one at velocity $v_1={348}\si{\nano\meter\per\second}$, and a second one at a velocity $v_2={5 109}\si{\nano\meter\per\second}$. By taking the difference between these two measurements, we cancelled all the velocity-independent forces  and from Eq.~\eqref{Taylor} we obtained $F_H$ measured at   $v =v_2-v_1={4 742}\si{\nano\meter\per\second}$ :
\begin{equation}
F_\text{total}(d,v'_2)-F_\text{total}(d,v'_1)=F_{H}(d,v)\,.
\label{eq:FH}
\end{equation}
\noindent Here, $v'_2$ and $v'_1$ are the relative velocities between the sphere and the sample, which are not exactly the piezo velocities $v_2$ and $v_1$ because the cantilever is deflected when the plate is moved towards the sphere. $v'_2$ and $v'_1$ were determined precisely measuring the deflection of the cantilever.

Measurements of the hydrodynamic force are presented in \cite{SuppMat}. Comparing the measured hydrodynamic force with the theoretical expression of Eq.~\eqref{Taylor}, we determined the separation distance upon contact $d_0=(31\pm2)\si{\nano\meter}$.

\textit{Electrostatic forces.--} Even if the surfaces are as clean as possible, there always remain electrostatic forces between them. First, an electrostatic potential difference $V_c$ still exists between clean, grounded, metallic surfaces owing to differences between the work functions of the materials used \cite{speake2003forces}. Second, electrostatic forces can remain due to the presence of trapped charges. In liquids, these trapped charges induce double-layer forces, due to the rearrangement of ions in solution, screening the electrostatic interactions. 

When  $d<<R$, the expression of the electrostatic force is \cite{laurent:tel-00576595}:
\begin{equation}
F_{e}=-\frac{\pi\epsilon_0 \epsilon_d R}{\lambda_D} {V_c ^2} \,\, \exp(-d/\lambda_D)
\label{eq_debye_force}
\end{equation}

The term $\frac{V_c^2}{d}$ is the contribution of the contact potential $V_c$ between the surfaces and the term $\exp(-d/\lambda_D)$ represents the double-layer force, screened over a distance $\lambda_D$  (the Debye length)\cite{butt}.

As there is no free charge in nitrogen, the Debye length is infinite and the electrostatic interaction is not screened, consequently there are no double-layer forces. In nitrogen, the contact potential was calibrated to $V_c=(87 \pm 2)\si{\milli \volt}$, and was compensated by an applied voltage difference during the measurement of the Casimir force.

\begin{figure*}[t]
\includegraphics[width=1\linewidth]{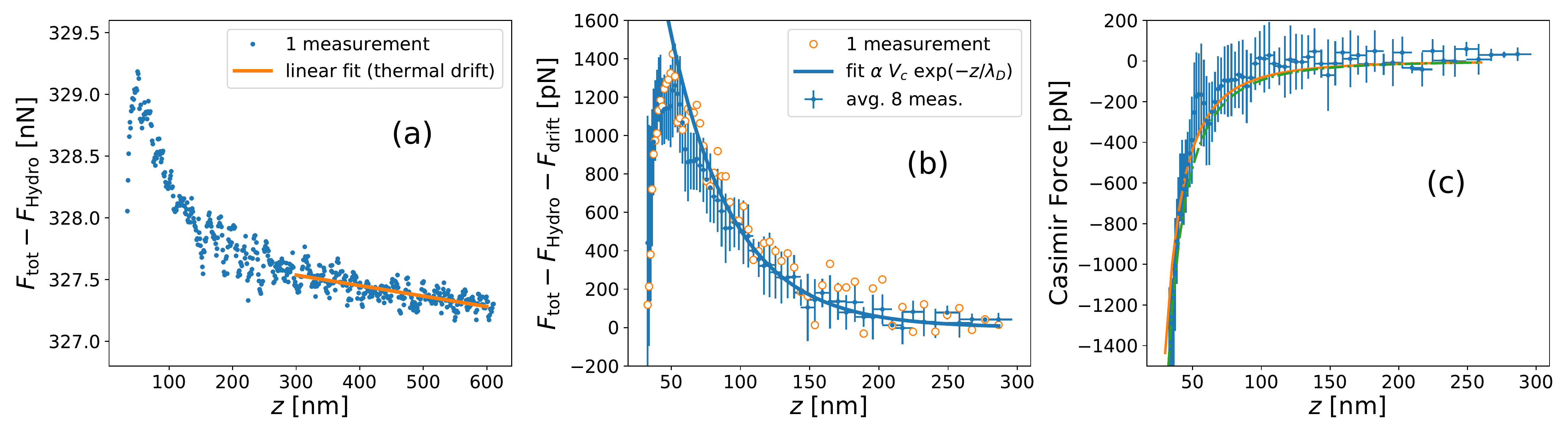}
\caption{\textbf{Measurement of the Casimir force in ethanol.} (a) Single force measurement (blue points) after subtracting the hydrodynamic force. The distance has been shifted by $d_{0}={31}\si{\nano\meter}$. The red line corresponds to the linear fit of the thermal drift in the range $300\si{\nano\meter}$ - $600\si{\nano\meter}$. (b) Single force measurement (red circles) and mean force measurement (blue points) (averaged over 8 measurements) of  $F_{\text{cas}}+F_{\text{Debye}}$ obtained after subtraction of the thermal drift and the hydrodynamic force. The mean measured force is fitted by an exponential in the range $d>120\si{\nano\meter}$. In this distance range, double-layer forces remove largely the Casimir force. The blue line corresponds to the exponential fit, yielding the Debye length $\lambda_D=(46 \pm 6)\si{\nano\meter}$ and the surface potential $\psi_{0_{adj}}=(63 \pm 13)\si{\milli\volt}$ through Eq.~\eqref{eq_debye_force}. (c) Measurement of the Casimir force between two Au-surfaces in a ethanol, where blue cross correspond to the mean measured force. The orange solid line corresponds to the Lifshitz theory where the dielectric function $\epsilon$ is evaluated from measured optical data of a real gold film \cite{PhysRevB_optical}; for the green dashed-dotted line, $\epsilon$ is evaluated using the handbook optical data \cite{palik1998handbook}. }
\label{force_casimir_ethanol}
\end{figure*}


In contrast, in ethanol, the contact potential is strongly screened by the ions constituting the Debye layer. Moreover, applying an electrostatic potential in a polar liquid can yield a transcient current \cite{FORT196846} and consequently, charges accumulate on surfaces, preventing us from applying the method suggested in \cite{PhysRevE.97.022611} to cancel the Debye force. Therefore, we simply subtracted the contribution of electrostatic forces from force measurements, after determining $\lambda_D=(46 \pm 6)\si{\nano\meter}$ and $V_c=(63 \pm 13)\si{\milli\volt}$ \cite{SuppMat}. In ethanol $V_c$ is lowered because the dissociation of molecules at the surface leads to the formation of a first very thin screening layer of a few $\si{\nano\meter}$.

\textit{Measurement of the Casimir force in nitrogen.--}\label{sec:meas_nitrogen} Static measurements of the Casimir force were carried out in a nitrogen atmosphere, between a gold-coated sphere and a gold-coated plate. We subtracted the vertical thermal drift by fitting linearly each force curve measurement between ${300}\si{\nano\meter}$ and ${1}\si{\micro\meter}$ and subtracting it from each measured curve.
All force curves were shifted in distance corresponding to the separation upon contact $d_{0}={31}\si{\nano\meter}$.

The measured Casimir force is shown on Fig.~\ref{force_casimir_nitrogen}(a) for separations ranging from ${90}\si{\nano\meter}$ to ${370}\si{\nano\meter}$, averaging $28$ independent measurements. For theoretical calculations, thermal corrections are negligible as the thermal energy $k_B T$ is too small to populate the mode of lowest energy $\hbar c/\lambda_T$, as the separation distance $d$ satisfies
\begin{equation}
d < 370\si{\nano\meter}  \ll  \lambda_T= \frac{\hbar c}{k_B T}\approx 7\,\si{\micro\meter} \,,\,\, \text{at}\,\,\, \SI{300}{\kelvin}\,.
\end{equation}
We compared our experimental result with theoretical predictions of the Casimir force, based on optical properties of Au taken from: $1$) handbook of tabulated data (green dashed-dotted line)\cite{Palik1997}, and $2$) measurements on Au samples presenting the same roughness and preparation conditions as ours (orange solid line)\cite{PhysRevB_optical}.  

The deviation from Lifshitz theory based on dielectric properties of real samples is less than $5\,\si{\pico\newton}$ at closest separations, while it reaches $10\,\si{\pico\newton}$ at closest separations for calculations based on data from handbook. As the signal to noise ratio degrades with increasing distance, the deviation of the measurements from Lifshitz theory increases at larger distances. However, they remain compatible within the error bars, including systematic and statistical errors (dominant at large distances). This demonstrates that surfaces must be carefully characterized for high precision measurements of the Casimir force. In order to make this argument more quantitative, we present the difference $(F_{exp}-F_{th})/F_{th}$ on Fig.~\ref{force_casimir_nitrogen}(b), showing the differences between the theoretical and experimental Casimir forces, for calculations based on data from handbook and calculations based on dielectric properties measured on films with the same morphology as our films.

\textit{Measurement of the Casimir force in ethanol.--} \label{sec:meas_ethanol} In a liquid, the scenario is richer than in a gas because of the presence of additional effects, namely the hydrodynamic force and the Debye screening of the electrostatic interactions. 

Measurements in ethanol were performed with the same apparatus, immediatly after the measurement in nitrogen, so that the contact are be the same, as explained previously. During the measurement of the Casimir force, the approach velocity was chosen in order to compromise between the hydrodynamic force $F_H$ we wanted to minimize and vertical drift, limiting the time of measurement. The results presented in this paper were obtained with ${100}\si{\nano\meter\per\second}$.

In order to average the data collected from consecutive runs, $8$ data sets were acquired. First, all force curves were shifted in distance corresponding to the separation distance upon contact $d_{0}={31}\si{\nano\meter}$. Second, the hydrodynamic force $F_H$ was subtracted from each measurement independently. As the $F_H$ dependence on distance is accurately known \cite{SuppMat}, it can be safely subtracted from the measured force. Third, to remove vertical thermal drift from force measurement, each force curve was fitted linearly between $300\si{\nano\meter}$ and $600\si{\nano\meter}$. Fig.~\ref{force_casimir_ethanol} (a) shows a single force measurement in ethanol after subtracting the hydrodynamic force. Blue points correspond to the raw measurement, and the red line corresponds to the linear fit of the thermal drift, which is then subtracted from each measurement. In Fig.~\ref{force_casimir_ethanol} (b), we represent both a single force measurement (blue) and the average measured force (red), after subtracting the thermal drift and the hydrodynamic force. The remaining force shows the presence of repulsive forces at separation distances larger than  ${60}\si{\nano\meter}$. These repulsive forces are attributed to the presence of ions in solution and on metallic surfaces. Then, the remaining curve corresponding to $F_{\text{cas}}+F_{\text{Debye}}$ was averaged over 8 measurements and the mean curve was fitted by an exponential function $A \exp(-d/\lambda_D)$ in the distance range $d>120\si{\nano\meter}$.
$A$ and $\lambda_D$ are adjustable parameters. In this distance range, we assume that the Casimir force is negligible in comparison to the double-layer force (as predictions of the Lifshitz theory indicate). We obtained a Debye length $\lambda_D=(46 \pm 6)\si{\nano\meter}$  consistent with measurements reported by \cite{van2009weak} and \cite{PhysRevA.78.032109}. The exponential fit is also used to determine the electrostatic potential at the gold surface $\psi_0$ from the expression of the double-layer force in a sphere-plane geometry: $F=4 \pi \epsilon \epsilon_0 R \psi_0^2 / \lambda_D \,\, e^{-d/\lambda_D}$. We evaluated the surface potential $\psi_{0_{adj}}=(63 \pm 13)\si{\milli\volt}$. After subtracting the measured double-layer force from each force measurement, the mesured Casimir force is obtained.

The measured Casimir force is presented on Fig.~\ref{force_casimir_ethanol}(b). The experimental data are compared to Lifshitz's theory for a gold sphere of radius $R={75}\si{\micro\meter}$ and a gold plate separated from a distance $d$ in ethanol. Finally the differences between the theoretical predictions and the measured data are plotted in Fig.~\ref{force_casimir_ethanol}(c). In spite of the rather large error bars we can  distinguish the two theoretical predictions: Casimir force measurements are in better agreement with Lifshitz theory based on optical properties of real Au films, presenting the same morphology as ours.

\textit{Conclusion.--}  In conclusion, we have presented measurements of the Casimir force performed both in gas (nitrogen) and liquid (ethanol) environments with the same apparatus and on the same spot of the sample. The force measurements yield experimental evidence of the importance of electrostatic effects in ethanol. These effects were properly measured and subtracted, in order to determine the genuine Casimir force. Furthermore, these measurements demonstrate that the Casimir force is sensitive to changes in the optical properties of gold at distances of less than $200\si{\nano \meter}$  mostly in the gas enviroment where the force is the strongest. Notably, to the best of our knowledge, this is the first time that this influence was measured experimentally at this range of separations. Our measurements are of significant interest given the fundamental implications of the Casimir force in the search of new hypothetical forces, technology applications of Casimir forces for micro/nano device actuation \cite{doi:10.1142/S021773230000102X,Genet}, and the very timely nature of our measurements \cite{PhysRevLett.120.040401} .

\vspace{1cm}

\section{APPENDIX}

\label{annexe}

This work has been supported by the ERC contract Outeflucoop. We thank Ir\'en\'ee Fr\'erot for helpful discussions.

\vspace{10pt}
$^*$ Present address of A.Le  Cunuder : \\ Department of Condensed Matter Physics,  \\University of Barcelona,\\ Marti i Franques, 1, \\ 080028 Barcelona, Spain

\section{Roughness analysis} \label{sec:rough}
 
 Because the Casimir force is sensitive to optical properties of gold films \cite{PhysRevB_optical}, we characterized carefully the topography of the surfaces. Indeed, it is commonly accepted that these properties can be taken from the handbooks tabulated data. In fact, optical properties of deposited films depend on the method of preparation. An interesting study \cite{PhysRevB_optical} reported a significant variation of $5-15\%$ in Casimir force calculations, due to changes in optical properties of Au films.
 
\begin{figure}[h!]
\centering
\includegraphics[width=0.9\linewidth]{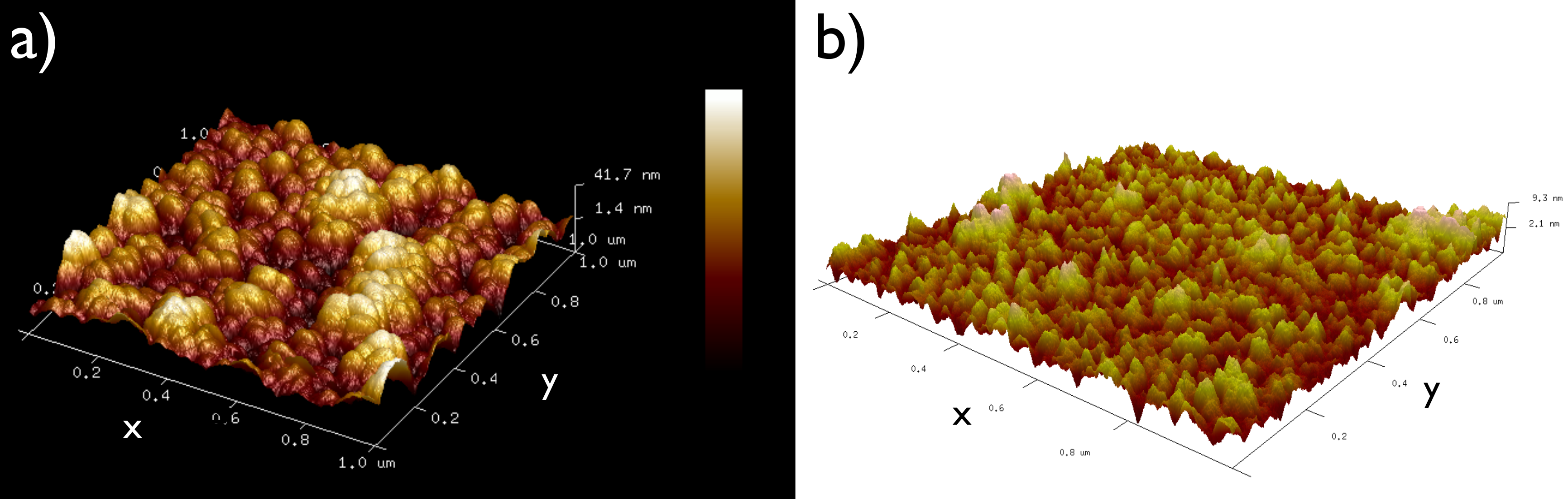}
\caption{\textbf{a} - AFM image of  $1\times1 \, \si{\micro\meter}^2$ sample of  the sphere surface  after the deposition of a {${100}\si{\nano\meter}$} thick gold film.   \textbf{b} AFM image ${1 \times 1}\,\si{\micro\meter}$ of the surface of the glass plate with  a {${100}\si{\nano\meter}$} thick gold film}
\label{fig:roughness}
\end{figure}

\begin{figure}[h!]
\centering
\includegraphics[scale=0.33]{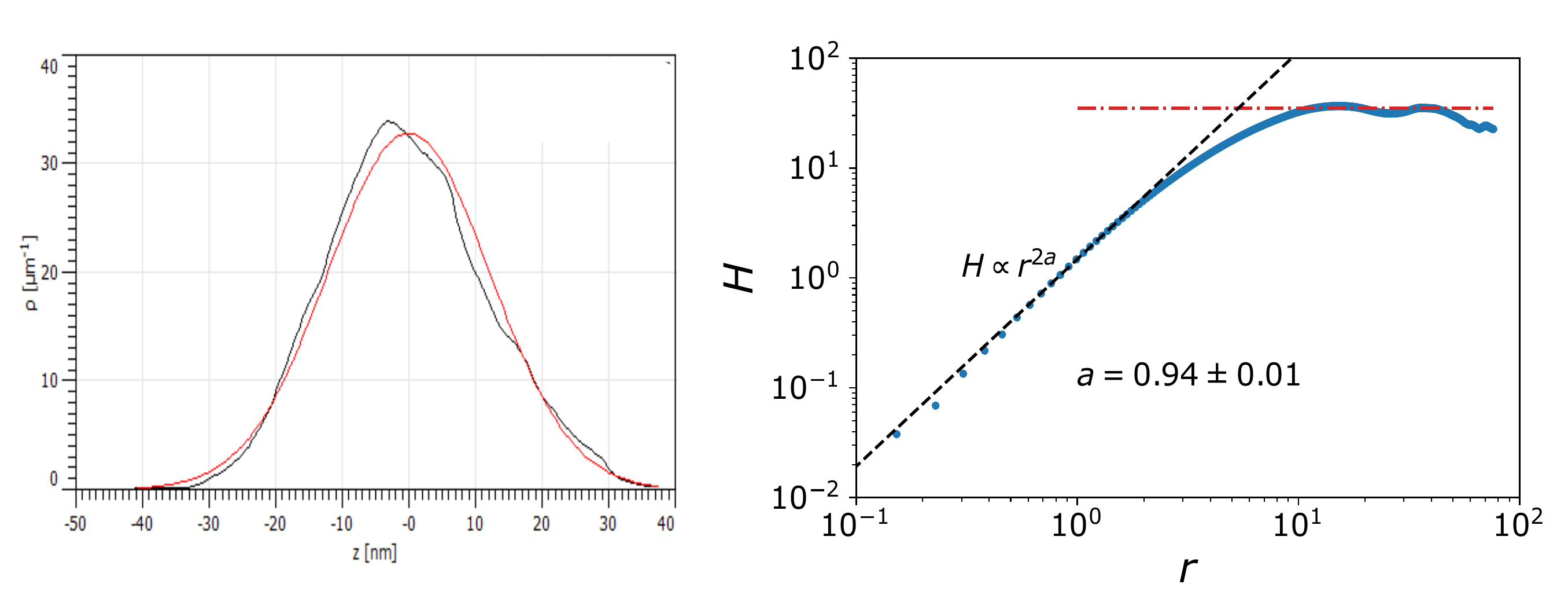} \  \

\caption{a) Height distribution of the sphere roughness
b) Height-height correlation function of the sphere surface in log-log scale. The height-height correlation function is defined as: $H(r)=<[h(r)-h(0)]^2>$, where $h(r)$ is the surface height. The roughness exponent $\alpha=0.940\pm 0.001$ is extracted from the slope of the linear fit and the correlation lenght $\xi=50\si{\nano\meter}$ is determined from the intersection between the linear fit and the saturation line.} 
\label{Height_distribution}
\end{figure}

After coating, the surface morphology of both the sphere and the plate was determined using a commercial AFM (Bruker). It is important to stress that these analysis were performed directly on the surfaces used in the measurement of the Casimir forces. The AFM images of a  $(1\times1 \, \si{\micro\meter}^2)$  sample of both surfaces are shown 
in fig.\ref{fig:roughness}. The roughness probability distribution of both surfaces are well approximated by a Gaussian. In fig.\ref{Height_distribution} a) we plot the probability distribution of the sphere whose rms roughness is $w_{sph}=({11.8 \pm 0.8})\si{\nano\meter}$, where the error takes into account the AFM accuracy and the statistical error based on a correlation length of about $50\si{\nano\meter}$, i.e. $400$ statistical independent points on the $(1\times1 \, \si{\micro\meter}^2)$ measured  surface. 

The correlation length, estimated from the Height-height correlation function plotted in fig.\ref{Height_distribution} b) is about $50\si{\nano\meter}$. The rms roughness of the plate is $w_{p}=({1.3 \pm 0.2})\si{\nano\meter}$.

This morphology analysis is then used in order to compare our force measurements with computations where optical properties are taken from real films, with a similar topography.

From morphology analysis, we also evaluated approximatively the separation upon contact $d_{0,rough}$. However, a surface of ${1 \times 1}\,\si{\micro\meter}$ is not large enough to determine precisely $d_0$. This analysis just helps us to check that we find a separation distance upon contact $d_{0,rough}$ of the same order of magnitude as $d_0$ obtained from hydrodynamic calibration. The AFM images \ref{fig:roughness} indicate that the sphere is much rougher than the plate. Consequently, the separation distance upon contact can be evaluated as: $d_{0,rough}=d_{0,sph}+w_{p}$, where $w_{p}$ is the rms roughness of the plate and $d_{0,sph}$ is the highest peak of the sphere within the contact area \cite{PhysRevB.93.085434}. One estimates that on a surface of  $1 \si{\micro\meter}^2$, there is in average only one asperity larger than $2.8 w_{sph}$ and less than one with height larger than $3  w_{sph}$, which is statistically coherent with the image of  fig.\ref{fig:roughness}a) and the tails of the distribution in fig.\ref{Height_distribution}. Thus, in a contact area of about $ 1 \, \si{\micro\meter}^2$ one expects to find $d_{0,sph}= 2.8 w_{sph}={(33.6 \pm 2.4)}\si{\nano\meter}$. Notice that this value is statistically significant because the area involved in the force measurements are of the order of $ 1 \, \si{\micro\meter}^2)$ at  $d<100\si{\nano\meter}$  (see ref.\cite{PhysRevB_distance}). For the plate, the rms roughness is $w_{p}=({1.3 \pm 0.2})\si{\nano\meter}$. Thus from the topography analysis we evaluate that the maximum is  $d_0 < ({34.9 \pm 2.4})\si{\nano\meter}$ on an area of $ 1 \, \si{\micro\meter}^2)$. This value is, within error bars, statistically coherent with the hydrodynamic calibration discussed in section 2. It is important to stress that the hydrodynamic calibration is performed directly on the surfaces used in the measurement of the Casimir forces. Indeed, calibration were performed in ethanol immediatly after measurements of the Casimir force. As the liquid was introduced very carrefully in the cell after the measurement in nitrogen and as the horizontal drift of the sample is negligible, the contact are is the same during both measurements and calibration. This assumption is further justified a posteriori: our experimental curves all superimpose on top of each others and on top of the theoretical curves after shifting the distance by the same value of $d_0$. In contrast the topography study is done on surfaces with the same statistical properties but not on the same position  of  the contact area used in the experiment. Thus we use for $d_o$ the value obtained from the hydrodynamic calibration, i.e.  $d_=({31 \pm 2})\si{\nano\meter}$.

\section{Hydrodynamic calibration} 
\label{hydro}

\begin{figure}[h]
\centering
\includegraphics[width=0.49\linewidth]{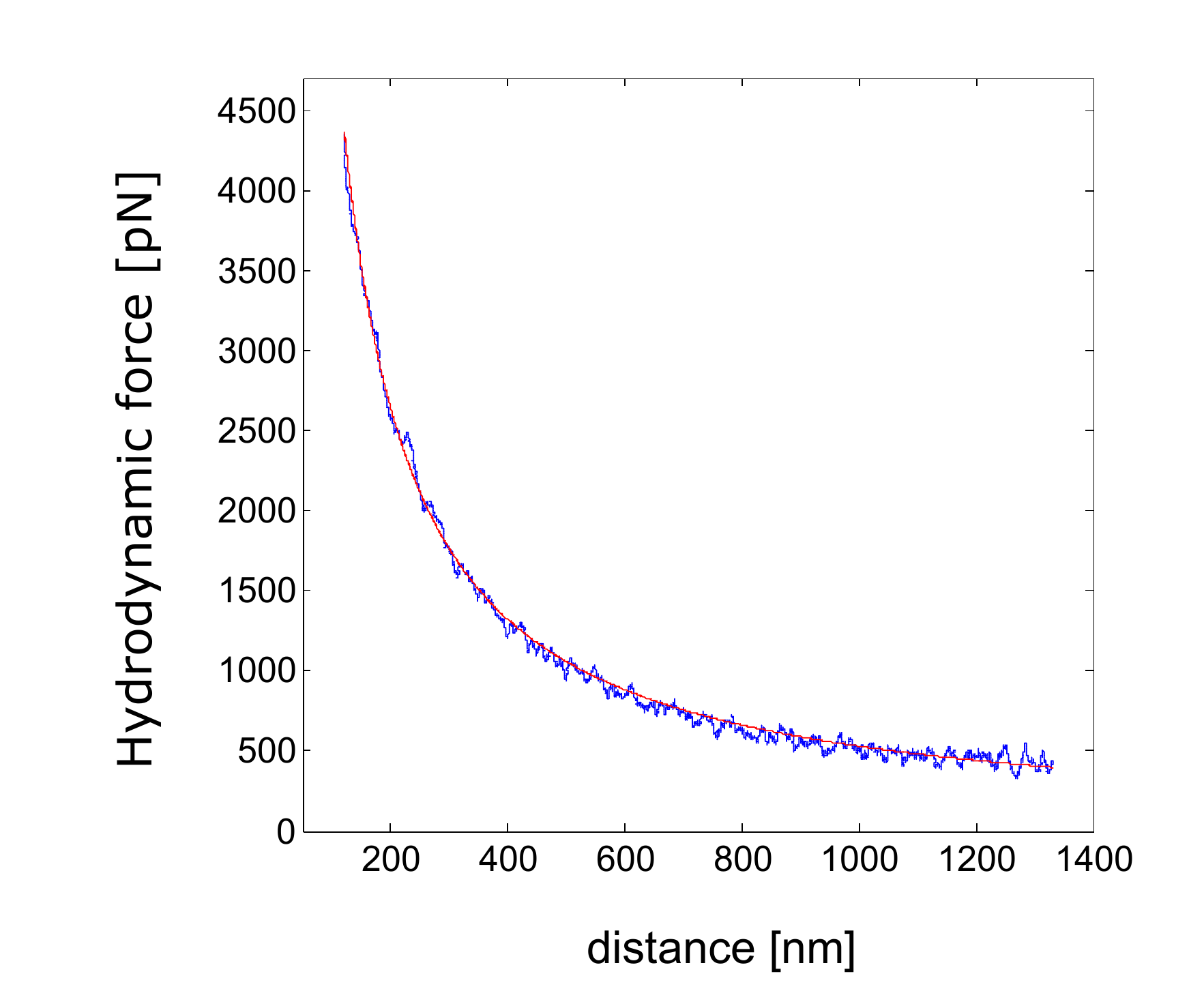}
\includegraphics[width=0.49\linewidth]{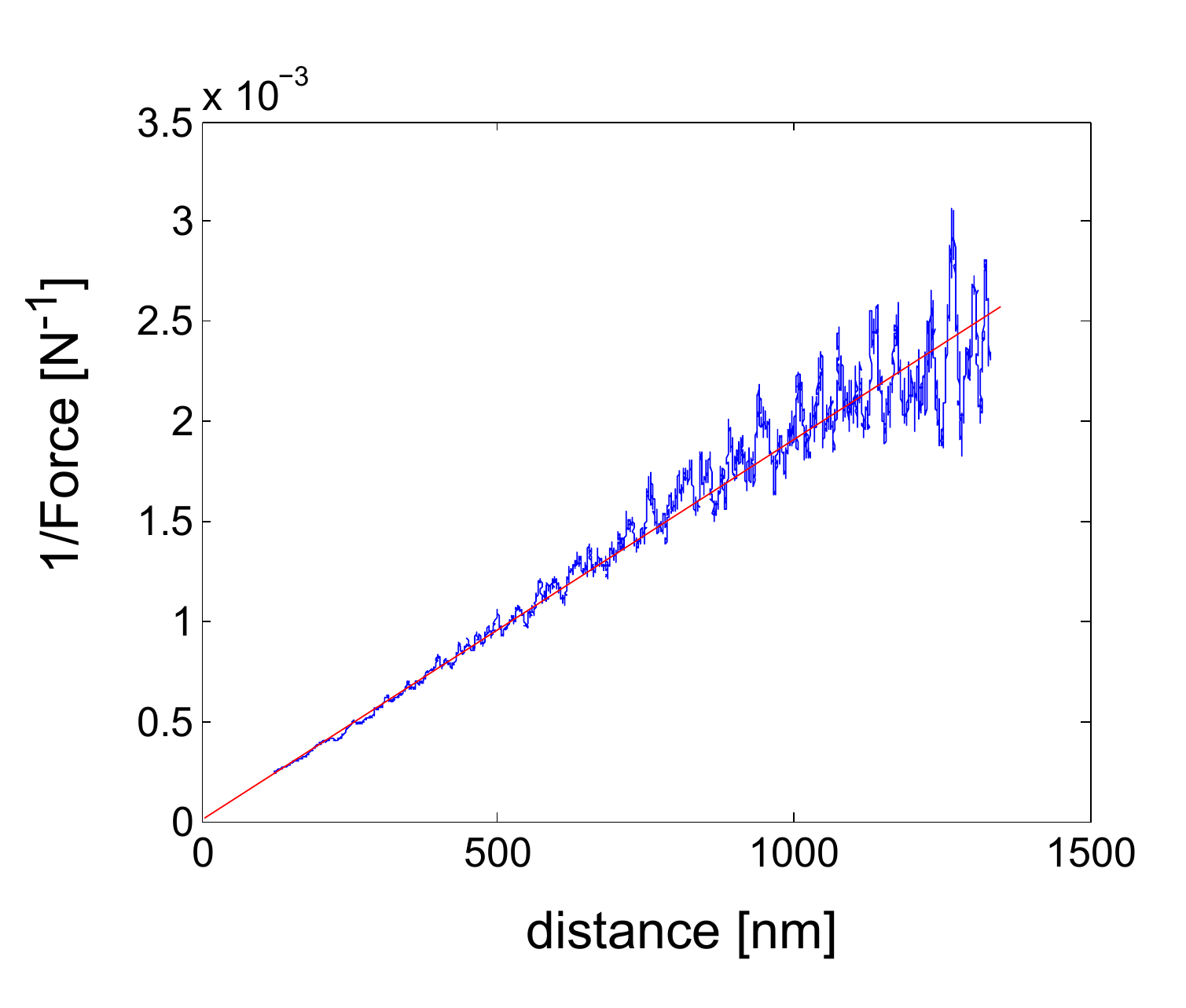}
\caption{a) Hydrodynamic force in ethanol measured at $v ={4.742}\si{\micro\meter\per\second}$.b) Inverse of the hydrodynamic force as a function of $d$.}
\label{Hydrodynamic force}
\end{figure}

The measured hydrodynamic force is plotted as a function of $d$ in fig.\ref{Hydrodynamic force} a) where it is compared to the theoretical force expressed in eq.(4) of main text. In the figure the measured values have been shifted horizontally by $d_o=31\si{\nano\meter}$ which corresponds to the separation distance upon contact.
In fig.\ref{Hydrodynamic force} b), the plot of the inverse of the force as a function of $d$ confirms that the value of $d_o$ is good since the curve crosses the origin. The slope $m$ of this  curve is in good agreement with the theoretical value $m_{th}$ determined from eq.(4) of main text. Specifically we find $1/m=1.98 \ 10 ^{-15} \si{\joule}$ and   $1/m_{th}=6 \pi \eta R^2 v=1.84 \ 10^{-15} \si{\joule}$, which are in agreement within the error bars of $d_o$, of $R$ and of $\eta$ . Thus, from hydrodynamic calibration, we get $d_o=(31\pm2)\si{\nano\meter}$.
 
\section{Calibration of the contact potential difference}
\label{subsubsec:contact_potential}
{An electrostatic potential difference $V_c$ exists between the sphere and the plate, even if the surfaces are coated with gold and they are both electrically grounded. Indeed, there can exist a large potential difference between clean, grounded, metallic surfaces owing to differences between the work functions of the materials used and the cables used to ground the metal surfaces \cite{speake2003forces}. A small potential difference around ten $\si{\milli\volt}$ is sufficient to overhelm the Casimir force so the contact potential difference has to be measured and the experiment has to be carried out with a compensating voltage present at all times.} 

{Following a procedure described in \cite{de2010halving}, we measure} the contact potential difference $V_c$ between the sphere and the plate by applying an oscillating potential $V={V_1} \cos{\omega_1 t}+{V_2}$ to the plate,  keeping  the sphere  grounded.

When  $d<<R$, the expression of the electrostatic force induced by the voltage potential difference ${V}$ and $V_c$ can be approximated by:
\begin{eqnarray}
F_{e}&=&-\frac{\pi\epsilon_0 \epsilon_d R}{d} {\left( {V_1}\cos(\omega_1 t)+V_2-V_c \right)^2} \notag \\
  &=&-\frac{\pi\epsilon_0 \epsilon_d R}{2d} \, [{V_1}\cos(2\omega_1 t)+ \notag \\ 
     &+&  4 {V_1} ({V_2}-V_c)\cos(\omega_1 t) +2 ({V_2}-V_c)^2+{V_1}^2 ] \label{eq:Fe_Vc}
\end{eqnarray}
Because of the existence of a contact potential difference $V_c$, the system oscillates both at $2\omega_1$ and at $\omega_1$. We determine $V_c$ adding a constant potential ${V_2}$ until the excitation at the frequency $\omega_1$ disappears. Indeed, when ${V_2}=V_c$, the system no longer oscillates at the frequency $\omega_1$ (see eq.\ref{eq:Fe_Vc}) .

We measure the contact potential $V_c$ as  a function of $d$ between ${1}\si{\micro\meter}$ and ${110}\si{\nano\meter}$. In practice, we move the plate towards the sphere by discrete displacements. At each separation distance $d_n$, we measure the potential $\overline{V_2}$ which minimizes the amplitude of the oscillation  at $\omega_1$.  \\
\begin{figure}
\centering
\includegraphics[width=0.8\linewidth]{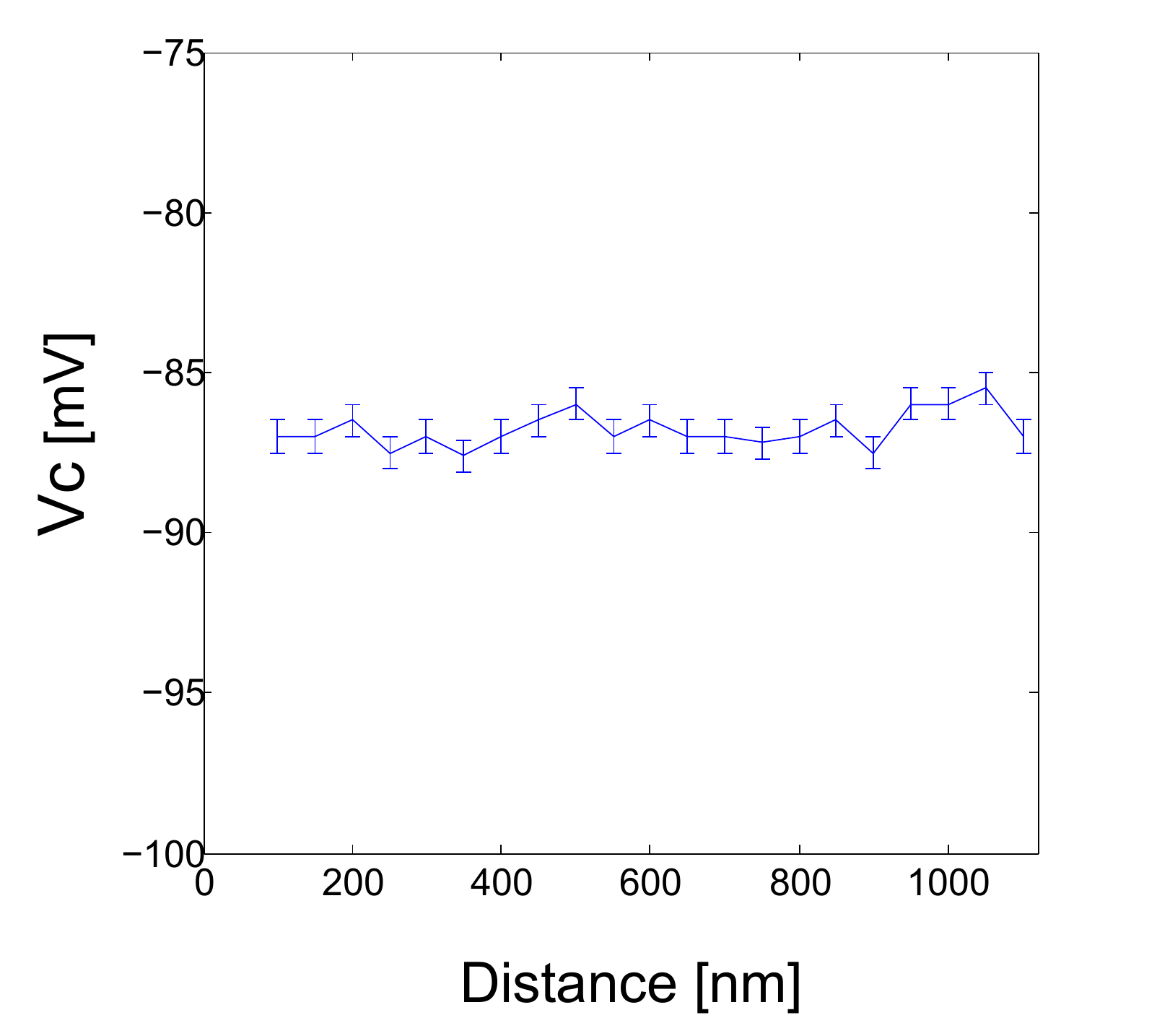}
\caption{Value of the potential ${V_2}$ which minimizes the excitation at frequency $\omega_1$. We measure a constant contact potential difference between ${110}\si{\nano\meter}$ et ${1.1}\si{\micro\meter}$.}
\label{fig:VC_versus_d}
\end{figure}
The result of the measurement is plotted in fig.\ref{fig:VC_versus_d}, where we see that in our experiment $V_c$ is constant as it is theoretically expected.

\bibliographystyle{unsrt}
\bibliography{bibliographie_article}

\end{document}